\documentclass[11pt,a4paper]{article}
\pdfoutput=1

\usepackage{jheppub}

\def\s{\sigma}

\def\br{\begin{eqnarray}}
\def\er{\end{eqnarray}}
\def\be{\begin{equation}}
\def\ee{\end{equation}}

\newcommand{\til}[1]{\tilde{#1}}

\title{Scale Factor Self-Dual Cosmological Models} 
\author[a]{U. Camara dS,} 
\author[a]{A. A. Lima} 
\author[a]{and G.M. Sotkov}
 \affiliation[a]{Universidade Federal do Esp\'irito Santo,\\ 
Esp\'irito Santo, Vit\'oria, Brazil} 
\emailAdd{ulyssescamara@gmail.com} 
\emailAdd{andrealves.fis@gmail.com} 
\emailAdd{gsotkov@gmail.com} 

\abstract{We implement a conformal time scale factor duality for  Friedmann-Robertson-Walker cosmological models, which is consistent with the weak energy condition.  
The requirement  for  self-duality  determines the  equations of state for a broad class of barotropic fluids. 
We study the example of a universe filled with two interacting  fluids, presenting an accelerated and a decelerated period, with manifest UV/IR duality. 
The  associated self-dual scalar field interaction  turns out to coincide with the ``radiation-like''  modified Chaplygin gas models.  We present an equivalent realization of them as  gauged K\"ahler sigma models (minimally coupled to gravity) with very specific and interrelated K\"ahler- and super-potentials. Their applications  in the description of hilltop inflation  and  also as  quintessence models for the late universe are discussed.}

\begin{document}

\maketitle

\setcounter{equation}{0}

\section{Introduction}
The  physical phenomena occurring at different periods of  the universe expansion are usually described by effective  cosmological models, representing gravity coupled to scalar fields with appropriately chosen interactions specific for the energy scales in consideration. The increasing precision of the recent astrophysical data \cite{2015arXiv150202114P,2015arXiv150201589P} allows to further select a 
restricted number of favored shapes of the corresponding  potentials \cite{Martin2013}, thus providing  important hints in the search of dynamical symmetry principles governing the universe evolution \cite{Hooft:2014kq,Kallosh:2013jk,Kallosh:2014rz,Ellis:2014lq, Hinterbichler:2011hl}.  At extremely high energies, where no preferred scales exist, the conformal or Weyl symmetries  completely determine  the corresponding effective gravity-matter actions \cite{Kallosh:2010uo,Kallosh:2002ph,Kachru:2003jt,Ferrara:2013fv,Ferrara:2010zp}. The description of lower densities and large scales phenomena, however, requires spontaneous or explicit breaking of the scale invariance  \cite{Hooft:2014kq,Kallosh:2013jk,Kallosh:2014rz,Hinterbichler:2011hl,Csaki:2014dk}. Hence an important problem to be addressed  concerns the nature of the additional symmetries that might be responsible for the selection  of  very special combinations of interaction terms giving rise to \emph{scale non-invariant} models with one of the favored inflaton or/and quintessence potentials \cite{Martin2013,Barreiro:2000mi,Copeland:2006if,Kamenshchik:2001ec,Caldwell:1998wq,Binetruy:1999qf,Brax:1999ee}. 

In order to highlight the power of discrete symmetries (and of the physical principles behind them) in the derivation of the matter interactions, we consider a particular class of cosmological models manifesting  discrete UV/IR  symmetry that relate the short-distances and early-times behavior to that at large-distances and late-times.  Although this symmetry should be broken at certain intermediate energy scales, it turns out to provide an effective description of the high energy features of the universe evolution in terms of the low energy ones.  A few examples of such UV/IR-self-duality can be realized in selected cosmological models  with two periods of decelerated/accelerated expansion, as  for example in  the $\Lambda$CDM model and its quintessence extensions \cite{Caldwell:1998wq,Barreiro:2000mi,Copeland:2006if,Kamenshchik:2001ec}. Our goal is to demonstrate that appropriate  transformations of the metric and energy density, which keep invariant the \emph{critical scale} ($L_{c}$) of the transition from radiation, matter or dark-matter (DM) dominated phase  to the dark-energy (DE) one, act as a specific \emph{discrete on-shell}  Weyl  symmetry of these models.  

Since the early and the late Universe is approximately homogeneous and isotropic, i.e. $ds^2=a^2(\eta)[-d\eta^2+ds_3^2(k)]$, it is natural to expect that the desired UV/IR  symmetries are related to the  \emph{scale factor duality} (SFD)  transformations \cite{Veneziano91scalefactor,Gasperini:2003sh,Chimento:2003il,Chimento:2008qq} of the corresponding FRW equations. Our main result consists in the implementation of the \emph{conformal time} scale factor duality in the way consistent with the weak energy condition. We also demonstrate that such SFD's, when combined with time reflections, can be promoted to manifest UV/IR symmetries for a broad class of cosmological models. The  requirement of self-duality allows us  to fix the form of the matter self-interactions as well.

\section{Conformal time Scale-Factor Duality} \label{SectCtSFD}

Let us introduce the following specific form  of the \emph{conformal time} SFD transformations of the  scale factor $a(\eta)$,  the energy density  $\rho(a)$ and pressure  $p(a)$  of the matter-fluid:
\begin{eqnarray} 
 \til a(\eta)=\frac{c^2_0}{a(\eta)},\quad\quad  \til {a}^2 \til{\rho}(\til a)= a^2 \rho(a),\quad \quad  \til {a}^2 [3\til {p}(\til a)+ \til{\rho}(\til a)] = -a^2 [3p(a)+\rho(a)]
 \label{sfdro}
\end{eqnarray}
As one can easily check, they  keep unchanged the form  of the FRW equations
\begin{eqnarray} 
 \kappa^2\rho=6\left(\frac{a'}{a^2}\right)^2+\frac{6k}{a^2},\quad\quad
 \rho'+3\frac{a'}{a}(\rho+p)=0,  \nonumber\\
 \kappa^2 p=2\left(\frac{a'}{a^2}\right)^2 -\frac{4a''}{a^3} -\frac{2k}{a^2},\quad \quad\kappa^2=16\pi G,\label{frw}
\end{eqnarray}
(with $ a'=\frac{da}{d\eta}$, etc.) \emph{independently} of the values of $ k=0,\pm1$, i.e. for spatially flat, open or closed universes.
By construction, they are mapping  expanding solutions  of the original model to contracting ones of its dual, $\til a'>0\mapsto{a}'<0$, and vice-versa. However, one can  use the invariance of  eqs. (\ref{frw}) under time reflections 
\be
\eta \mapsto \tilde \eta \equiv 2\eta_c-\eta \ ,\quad  a(\eta) \mapsto a(\tilde \eta),\label{reflect}
\ee
which in composition with the scale factor duality (\ref{sfdro}),
\br 
 \til {a}(\tilde \eta)=\frac{c^2_0}{a(\eta)} ,
\label{tau}
\er 
allows to relate two dual solutions both expanding or contracting, i.e. $a'>0 \mapsto \bar {a}'>0$, etc.  This particular form ensures that the critical time $\eta_c$, corresponding to the  fixed point $\til {a}(\eta_c)=a(\eta_c)$, with $c_0^2=a(\eta_c)^2$, remains invariant.

We next  consider a few  consequences of eqs. (\ref{sfdro}), representing the main features of the corresponding dual pairs of cosmological models in the case of fluids with generic equation of state (EoS),  $p=\omega(\rho)\rho$ and $\til p=\til{\omega}(\til {\rho})\til{\rho}$.

\emph{Matter/DE duality}: Since  the deceleration parameters $q$ and $\til{q}$  have always opposite signs $q(\omega)=-\til{q}(\til{\omega})$, i.e.
\begin{eqnarray} 
& \omega(\rho)+\til{\omega}(\til{\rho})= -\frac{2}{3},\quad q(\omega)=\frac{1}{2}[1+3\omega(\rho)][1+k\big(\frac{a}{a'} \big)^2],\label{qq}
\end{eqnarray}
we realize that the entire radiation, matter or DM dominated  phases of decelerated  expansion are always mapped into the corresponding DE dominated accelerated expansion periods in the dual model. The fact that the transformations (\ref{sfdro}) give rise to  such a \emph{matter/DE  duality}  becomes evident in  the case of perfect fluids with $\omega=const$. Then all the DE fluids  $-1\leq\omega\leq-\frac{1}{3}$ are transformed into ``matter-like'' ones\footnote{We shall often refer to such fluids with $\omega\in(-\frac{1}{3} , 1)$, which include dust, radiation, etc., loosely as ``matter fluids'', while the fluids with $\omega \in (-1,-\frac{1}{3})$, as usually are called ``dark energy''.}, with $-\frac{1}{3}\leq \omega\leq \frac{1}{3}$, and vice-versa. 
We have thus examples of pairs of dual Friedmann universes -- e.g. the de Sitter universe ($\omega = -1$)  is dual to the radiation filled universe ($\omega = 1/3$),  corresponding to the well known inverse proportionality relation between their scale factors, 
\br
a^2_{rad}(\eta) \sim \eta^2 \; ; \quad {\text{and}} \quad a_{dS}^2(\eta)\sim 1 /\eta^2 . \nonumber
\er
A dust dominated universe ($\omega = 0$), with $\rho_{\omega = 0} \sim 1 / a^3$, is dual to  cosmic domain-walls,  $\rho_{\omega = -2/3} \sim 1 / a$. The cosmic string gas, with $\omega_{str}=-\frac{1}{3}$, is dual to itself. 
Notice that outside the range of $-1 \leq \omega \leq \frac{1}{3}$, the duality allows us to describe certain ``phantom DE fluids'', with $- \frac{5}{3 }\leq \omega < -1$, in terms of ordinary matter  fluids with $\frac{1}{3} <\omega \leq 1$.

\emph{Weak energy condition's consistency}:  An important property of the conformal time scale factor duality (\ref{sfdro}) is that it preserves the weak energy condition (WEC) for a broad variety of fluids. That is, if we require that the WEC ($p+\rho\geq 0$ and $\rho\geq 0$) is satisfied by \emph{both} dual fluids,  we find that the EoS parameters $\omega(\rho)$ and $\tilde \omega(\tilde \rho)$ are restricted to the interval $[-1,1/3]$. This  condition is not only fulfilled by all perfect fluids with constant $\omega$  such as dark energy, cold (dark) matter and radiation, but also by countless examples of  dual models involving  more complicated interacting multi-component fluids.

\emph{UV/IR duality}: The scale factor inversions, independently of their cosmic \cite{Veneziano91scalefactor, Gasperini:2003sh, Chimento:2003il, Chimento:2008qq} or conformal time realizations, are always mapping the short distance scales (at certain fixed time $\eta$) in the original model to the large scales (at the same $\eta$) in the dual model and vice-versa. In the case of conformal time SFD's (\ref{sfdro}), however, these  relations are enhanced  by an important new UV/IR ingredient. Namely, for WEC consistent fluids high energy densities $\rho$ and high scalar curvatures $R$  are always  transformed to lower ones in the dual models.
Perfect fluids, with $\rho_{\omega}=D_{\omega} a^{-3(1+\omega)}$  ($-1< \omega < \frac{1}{3}$) 
provide  the simplest example of such UV/IR duality: their duals are again perfect fluids with $\til{\omega}=-\frac{2}{3}-\omega$ and $\til {\rho}= \rho^{\frac{3\omega-1}{3(\omega+1)}}$.  
The   curvature transformation $\til {R}(R)$ turns out to be similar to  that of $\til {\rho}(\rho)$, since as a consequence of eqs.(\ref{frw}) we have  $R=\frac{\kappa^2}{2}(1-3\omega)\rho$.
As expected, for the self-dual fluid, $\omega_{str}=-\frac{1}{3}$, we recover the proper inversion rule $\til {R}_{str}=\kappa^4/R_{str}$. Similar high/low densities and curvatures transformations  take place for certain cosmological models involving two-component interacting  $\omega (\rho) \neq const$ fluids considered  in Sect.\ref{SectSDFluids} below.

\emph{Duality of event and particle horizons}:  As a consequence of the scale factor inversion followed by time reflection (\ref{tau}), the sign of the Hubble factor $H(a(\eta))=\frac{a'}{a^2}$ is kept unchanged and  its transformation law is given by  $\tilde a \tilde H = a H$ (cf. eqs. (\ref{sfdro})). This property  allows us to  demonstrate that scale factor duality maps a particle horizon in the original universe into a cosmological event horizon of its dual, and vice-versa. Take, for simplicity, flat, singular universes with scale factor in the range $(0,\infty)$. Then $r_e = \int_{a}^{\infty} \frac{da}{H a^2}$ and $r_p =  \int_{0}^a \frac{da}{H a^2}$ are the comoving radii of event and particle horizons, respectively, and one can easily derive their duality transformations:
\br
\tilde r_p = \int_{0}^{\tilde a} \frac{1}{\tilde H \tilde a} \; \frac{d \tilde a}{\tilde a} = - \int_{a(\tilde a = 0)}^{a(\tilde a)} \frac{1}{H a} \frac{da}{a} =  \int^{\infty}_{a} \frac{1}{H a} \frac{da}{a} = r_e,\label{dhor}
\er
while for the physical radii $L_e=a r_e$ and $L_p=a r_p$  we obtain:
\br 
  \frac{L_p}{a}=\frac{\til{L}_e}{\til{a}} .\label{dhors}
\er
For $k \neq 0$ the same reasoning applies. Thus the dual of a universe possessing a particle horizon necessarily presents an event horizon, and vice-versa. This relates their causal structure: if the original universe has a space-like past infinity, its dual will have a space-like future infinity, while a null past infinity is dual to a null future infinity. Such is the case for matter fluids (presenting space-like past infinity and particle horizon) and DE fluids (presenting space-like future infinity and event horizons), in accordance with (\ref{qq}).

\vspace{0.5cm}

The scale factor duality, originally  introduced as  an extension of the superstring  T-duality for time dependent cosmological  backgrounds of dilaton gravity \cite{Veneziano91scalefactor,Gasperini:2003sh},  admits a few different cosmic and conformal time realizations, both in Einstein and in dilaton-axion gravities. It is known to be an essencial ingredient of the alternative  models of dilaton driven pre-big-bang  inflation. Its Einstein gravity cosmic time implementation  has been used for the description of  ``phantom  models''  of dark energy  \cite{Chimento:2003il,Chimento:2008qq}.  Although our motivations to study the cosmological implications of the \emph{conformal time}  scale factor duality  mainly concerns its applications as an UV/IR symmetry  of the two phases of the \emph{post-inflationary} evolution of the universe,  we find worthwhile to briefly stand out the diferences and the similarities  between the properties of these three  families of scale factor dual cosmological models.

\emph{Cosmic time SFD in Einstein gravity:}  The  cosmological models obtained by imposing  the requirement of  scale factor duality in \emph{cosmic time},
$ds^2=-dt^2+a^2(t)ds_3^2(k=0)$, 
\br
 \til {a}(t)=\frac{c^2_0}{a(t)},\quad\quad \til {\rho}_{cos}=\rho_{cos},\;\; \;\;\til {p}_{cos}= - p_{cos}- 2\rho_{cos},\;\; \;\;\omega_{cos}+\til{\omega}_{cos}= - 2 , \label{chimentdual}
\er
studied in refs.\cite{Chimento:2003il,Chimento:2008qq}, manifest quite different features from the  \emph{conformal time}  dual models considered in the present paper. The fact that the scale factor transformations leave unchanged the corresponding energy densities is the origin of an unavoidable presence of  ``phantoms'' ($\til {\omega}_{cos} < -1$), as well as of the violation of  the weak energy condition  and, as a consequence, of their non-conventional UV/IR behavior. 

\begin{figure}[ht] 
\centering
\includegraphics[scale=0.5]{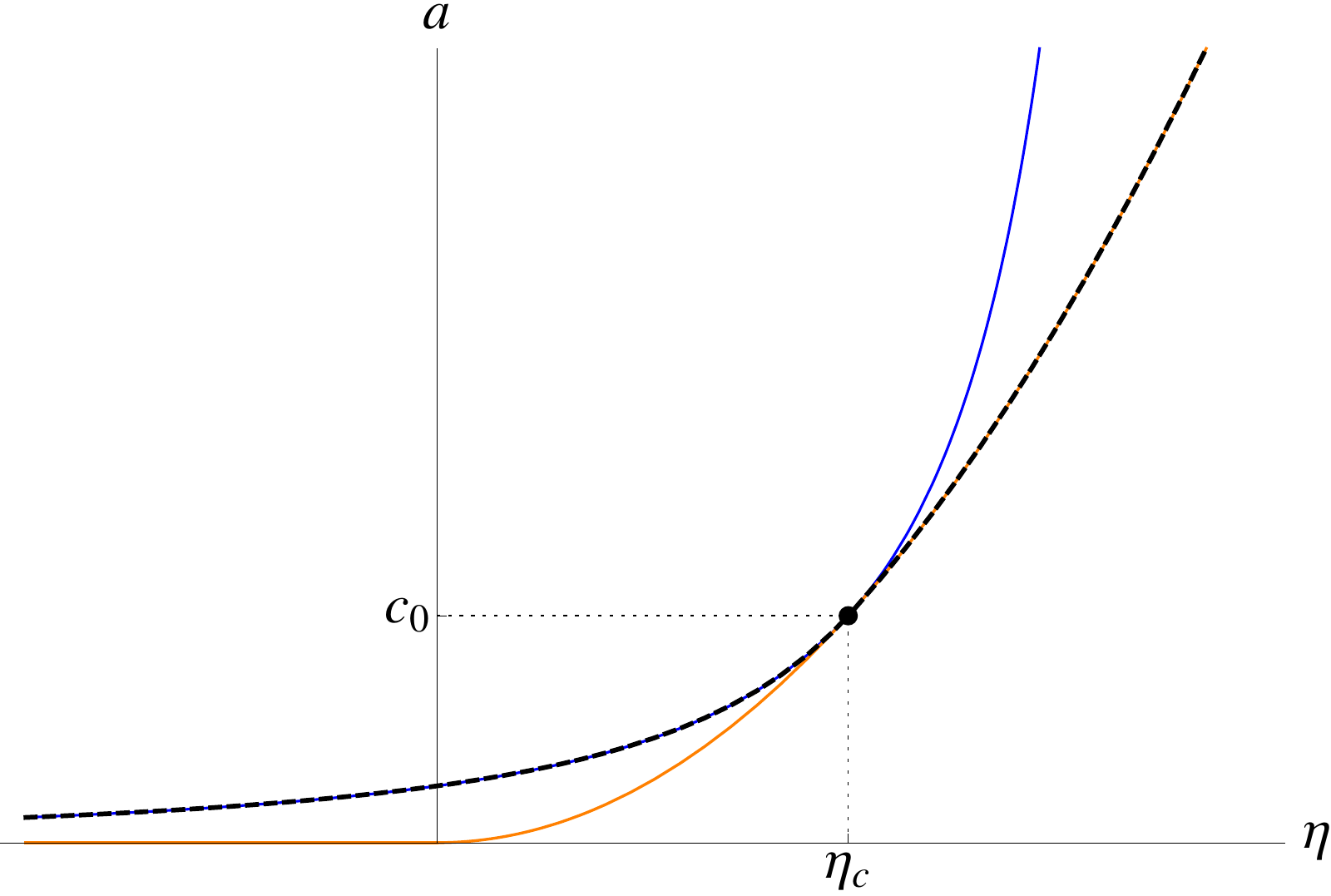}
\caption{Self-dual solution obtained from the composition of two dual perfect fluids. The blue line represents $a_{\omega}(2\eta_c - \eta)$ for $\omega = -\tfrac{2}{3}$ (cosmic domain walls); the orange line represents its dual $a_{\tilde\omega}(\eta)$ for $\tilde\omega = 0$ (dust). The black dashed line is the continuous, piecewise function $a(\eta)$, with $a(\eta) = a_{\omega}(2\eta_c - \eta)$ for $\eta \leq \eta_c$, and $a(\eta) = a_{\tilde\omega}(\eta)$ for $\eta \geq \eta_c$.  
}
 \label{GLUE}
\end{figure}

\emph{Dilaton gravity self-dual  models:}  The original cosmic time scale factor duality  transformations \cite{Veneziano91scalefactor,Gasperini:2003sh}, 
\br
  a(-t)=\frac{1}{a(t)},\quad \til{\varphi} = \varphi -6\ln a,  \quad \til {\rho}_{dg}=a^6 \rho_{dg},\;\; \;\;\til {p}_{dg}= -a^6 p_{dg},\;\; \;\;\omega_{dg}+\til{\omega}_{dg}=0 , \label{dgdual}
\er
act as  a discrete $Z_2\otimes Z_2$  symmetry of the homogeneous and isotropic  solutions of  dilaton gravity\footnote{Realized in the string (i.e. Jordan) frame.} with a dilaton field $\varphi(t)$ coupled to specific (dual) matter fluids.
As usual, the dilaton  has a vanishing potential and the standard  interaction  with  perfect fluids of density $\rho_{dg}$ and pressure $p_{dg}$, is assumed. 

Despite the evident differences  in the realizations of the corresponding cosmic and conformal time scale factor duality (\ref{sfdro}), (\ref{tau}) and (\ref{dgdual}),  the problems addressed  are quite similar: the construction of physically consistent self-dual solutions, presenting two phases (accelerated and decelerated) that are joined at the fixed points of time-reflection, $t_c$ or $ \eta_c$.  However, the presence or the absence of a dilaton allows for non-equivalent choices of the  instants  of  acceleration/deceleration transition, and of different realizations  of the dual or self-dual fluids, resulting in cosmological solutions with qualitatively different  properties. 

The simplest pre-big-bang  cosmological solutions of dilaton gravity are obtained by
 considering  a  pair of perfect fluids, dual with respect to (\ref{dgdual}),
 one with  accelerated  and the other with decelerated expansions, ``glued'' together at $t=0$. 
The result is a self-dual cosmological model with a pre-big-bang inflationary phase in $t<0$ driven by a nontrivial dilaton $\varphi(t)$, and followed by the standard $\Lambda$CDM matter/radiation dominated period in $t>0$, with a constant dilaton. These models entail a  curvature singularity and an \emph{infinite} jump of the dilaton at the ``big-bang'' in $t=0$ \cite{Veneziano91scalefactor,Gasperini:2003sh}.
 
The  conformal time Einstein gravity cosmological models obtained from (\ref{sfdro}) admit similar, but less singular and even  completely non-singular, self-dual solutions, now in the Einstein frame. 
Indeed, a similar gluing procedure may be applied, e.g., to dual pairs of perfect fluids with $\omega\in(-1,1/3)$, related by (\ref{sfdro}). We join the scale factors at the point of time reflection $\eta_c$, when the dual scales coincide: $a(\eta_c) = \tilde a(\eta_c) = c_0$.
 The result, shown in fig.\ref{GLUE}, represents a simple two-phases universe evolution starting with a ``big-bang'' singularity (at $\eta=-\infty$), undergoing a phase of ``inflationary'' (i.e. accelerated) expansion up to $\eta_c$, when  the deceleration period (driven by the dual  fluid) begins.%
 \footnote{This procedure always yields  \emph{two diffrent} self-dual universes: the other one has a big-bang at $\eta = 0$, followed by a \emph{decelerated} expansion up to $\eta = \eta_c$, when it becomes  accelerated until  its end at $\eta = 2\eta_c$.}
This composite solution  is ``almost smooth'', with only a \emph{finite} jump of the deceleration parameter $q(a)$ and of the curvature at the joint at $\eta_c$. One can also construct  physically more interesting universes presenting a \emph{completely smooth} transition between the accelerated and decelerated phases. These models are based on  a particular family of two interacting \emph{self-dual} fluids with SFD invariant EoS derived  in Sect.\ref{SectSDFluids}, and further studied  in Sect.\ref{SectUVIR} below.


Our final remark concerns the common UV/IR features of the considered  Einstein and dilaton gravity  dual  cosmological models.  One can easily verify that  eq.(\ref{dgdual}) give rise to a ``density inversion law''  $ \til {\rho}_{dg}\propto \rho_{dg}^{\frac{\omega-1}{\omega+1}}$ for  a pair of dual perfect fluids which shows that,  similarly to the Einstein gravity conformal time dual models, also in dilaton gravity the  high energy densities are mapped by scale factor duality (\ref{dgdual}) to low densities and vice versa.


\section{Self-Dual Fluids} \label{SectSDFluids}

The properties established above provide a clear indication on how to convert the duality transformations (\ref{sfdro}) between \emph{pairs} of dual cosmological models into an effective $Z_2\times Z_2$ symmetry of a \emph{single} FRW solution by requiring its  \emph{self-duality}:
\be 
\tilde{a}(\eta)=a(\tilde{\eta}),\quad\quad \til{\rho}(\til a)= \rho(\til a),\quad\quad  \til{p}(\til a)= p(\til a) ,
\label{sfsd1}
\ee
with $\tilde \eta = 2 \eta_c - \eta$ given by the time reflection (\ref{reflect}). 
These constraints allow us to find the EoS for families of self-dual fluids, with $\til{\omega}(\til {\rho})=\omega(\til{\rho})$, as solutions to the functional  equations
\br
\frac{c_0^2}{a(\eta)}=a(2\eta_c - \eta)  , \quad \rho( {\Omega} a)= {\Omega}^{-2}\rho(a),  \quad\omega(\rho)+ \omega({\Omega}^{-2}\rho)= -\frac{2}{3},
 \quad   \Omega\equiv\frac{c_0^2}{a^2} .
\label{homog} 
\er
As suggested by the proper definition of the pairs of dual fluids with $\omega=const$, 
one possible solution is the sum of their densities,
\be
\rho_{\omega}^{sd}(a)= B a^{3\omega-1} + \tilde B a^{-3(1+\omega)}, \quad\quad \tilde B = B c_0^{2(1+3\omega)}\nonumber .
\ee
With the addition of other pairs of dual fluids (and/or the self-dual fluid $\rho_c=\Omega_c a^{-2}$), one may construct more realistic models, as for example 
\be  
\rho_{{\rm{\Lambda CDM}}}^{sd}=\Omega_{\Lambda}+\frac{\Omega_{rad}}{a^4}+\frac{\Omega_{dw}}{a}+\frac{\Omega_{dust}}{a^3}+\frac{\Omega_{c}}{a^2} . \label{lcdm}
\ee
This is nothing but the standard $\Lambda$CDM model enhanced by a new term, $\rho_{dw}=\Omega_{dw}/a$, representing  cosmic domain walls with  $\omega_{dw}=-2/3$. The fraction of the critical energy density due to this contribution is set fixed by the constraint of self-duality as 
\br
\Omega_{rad}/\Omega_{\Lambda}=(\Omega_{dust}/\Omega_{dw} )^2=c_0^4. \label{relatvieOmegafix}
\er 

The most general solution of the eqs. (\ref{homog}) for the energy density of a two-component self-dual fluid takes the form:
\be 
\rho=\Big(B a^{-3(1+\beta)}+ D a^{-3(1+\gamma)}\Big)^{\frac{4}{3(\gamma+\beta+2)}}, \label{twogen}
\ee 
with $D=B c_0^{3(\gamma-\beta)}>0,\  (\gamma+1)(\beta+1)\geq 0$. Its asymptotic behavior for small and large values of the scale factor is described by a  family of  dual pairs of perfect fluids with EoS parameters
\begin{eqnarray} 
\omega_{\gamma}=\frac{\gamma-3\beta -2}{3(\gamma+\beta +2)}, \quad\quad  \omega_{\beta}=\frac{\beta-3\gamma -2}{3(\gamma+\beta +2)} ,\quad\quad \omega_{\gamma}+\omega_{\beta}= -\frac{2}{3}.
\label{sfluids}
\end{eqnarray}
Notice that all the self-dual fluids (\ref{twogen}) satisfy  the same  ``boundary'' conditions: $\rho \to \{0 \ {\text{or}} \ \infty\}$ when $a\to\{\infty \ {\text{or}} \ 0\}$, respectively. 
The exception is for $\beta=-1$, with an arbitrary $\gamma\equiv\frac{4}{3} \delta - 1$. Then one of the asymptotic limits leads to a finite density, $\rho\to\rho_{dS}\equiv B^{\frac{1}{\delta}}$, and is a de Sitter universe with EoS parameter $\omega_{\beta}=-1$. On the opposite  limit we find radiation, with $\omega_{\gamma}=1/3$. The self-dual fluid in this case is actually a modified Chaplygin gas  \cite{Debnath:2004cq,Benaoum:hh}, with a rather simple EoS
\begin{eqnarray} 
& \rho =\Big(B + D a^{-4\delta}\Big)^{\frac{1}{\delta}} , \ \ \ p=\frac{1}{3} \rho -\frac{4}{3} B \rho^{1-\delta}.
\label{rChap}
\end{eqnarray}

The same approach  can be used in the construction of more general self-dual energy densities, as for example, 
\be 
\rho = \sum_\alpha \left( C_\alpha + E_\alpha a^{-r_\alpha} \right)^{p_\alpha} \left( B_\alpha + D_\alpha a^{-s_\alpha}\right)^{q_\alpha}; \quad\quad r_\alpha p_\alpha + s_\alpha q_\alpha = 4
,\label{MultiChap}
\ee
representing a composition of multiple different Chapligyn gas-like constituents. Each of the terms in the sum are separately self-dual given that their coefficients are related by
\br
E_\alpha = c_0^{r_\alpha} \ C_\alpha \ ; \quad D_\alpha = c_0^{s_\alpha} \ B_\alpha \ .\label{sdcoeffrelamutlichap}
\er
Such combinations allow for richer phenomenological applications of self-dual models -- but, on the other hand, they introduce many new free parameters and thus call for further physical requirements  in order to select one (or a few) among the allowed densities.

\vspace{0.5cm}

In the following, we shall restrict ourselves  to  the study  of  the simplest two-fluids family of  self-dual  models (\ref{rChap}).  They permit a rather explicit analytic description, which makes evident the UV/IR nature of the  corresponding conformal time SFD transformations (\ref{sfsd1}) and (\ref{homog}).  As we have already mentioned, the scale factor inversions  always map the short distance scales  into large distance ones in the dual model, and vice-versa \cite{Veneziano91scalefactor,Gasperini:2003sh}.  This universal property  is now complemented by the remarkably simple  energy density and scalar curvature ``inversion laws'':
\br
\til {\rho}^{\delta} - \rho_{dS}^{\delta}=\frac{\rho_{dS}^{2\delta}}{\rho^{\delta}-\rho_{dS}^{\delta}} ,\quad\quad\quad  \til {R}^{\tau} - R_{dS}^{\tau}=\frac{R_{dS}^{2\tau}}{R^{\tau}-R_{dS}^{\tau}},
\label{roro}
\er 
with $\tau=\frac{\delta}{1-\delta}$.
They are obtained by substitution of  $a(\rho)$ from eq.(\ref{rChap}) into the $\rho$ transformation law (\ref{homog}), i.e.  in $\til{\rho}= \frac{a^4}{c_0^4} \rho $, and by taking into account 
the relation  between curvatures and densities for the modified Chaplygin gas  (\ref{rChap}):
\br
& R=R_{dS} \left( \frac{\rho}{\rho_{dS}} \right)^{1-\delta} , \quad\quad\quad R_{dS}=2\Lambda_{dS}=\frac{12}{L_{dS}^2}.\label{curvtransinvert}
\er
Hence, for the considered self-dual  models with $\delta<1$, high densities and curvatures are  transformed into small ones and vice-versa. The critical density $\rho_c \equiv \rho(\eta_c)$, which marks  the transition between decelerated and accelerated phases, is defined as  a fixed point $\tilde \rho_c = \rho_c$ of the above transformation (\ref{roro}).
With the help of  the curvature radius $L^2 \equiv 12 / R$  transformation law 
\br
\til {L}^{\frac{2\delta} {1-\delta}} +  L^{\frac{2\delta} {1-\delta}} = L_{dS}^{\frac{2\delta} {1-\delta}} , \label{lcurv}
\er
one can also calculate the ``critical curvature scale''  $L_c = \tilde L_c$  in terms of the (asymptotic) de Sitter radius: $L_c=2^{\frac{\delta-1}{2\delta}} L_{dS}$.

\section{Reconstruction of the self-dual matter potentials}

The requirements for scale factor duality  of the FRW equations (\ref{frw}) can be also  realized in an equivalent form  by replacing the  matter  fluid  with  one scalar  field $\s(\eta)$  with potential $V(\s)$, as usual: 
$$\rho=\frac{1}{2} (\s'/a)^2+V, \quad\quad\quad p=\frac{1}{2}(\s'/a)^2-V, $$
 where  $ \s'=\frac{d\s}{d\eta}$. An efficient method for the explicit construction of  self-dual scalar field potentials is provided by  defining a   superpotential  ${\cal W}(\s)$  \cite {Cvetic:1995jo,Cvetic:1997fx,Townsend:2007ta,Skenderis:2007yb}  and the related BPS-like first order equations:
\begin{eqnarray}
a'^2=\frac{a^2}{4}(\kappa^2 a^2{\cal W}^2-4k),\quad\quad  \s'^2= - \frac{a^3}{2} \frac{d{\cal W}^2}{da},\quad\quad  V(a)=\frac{3}{2}{\cal W}^2 + \frac{a}{4}\frac{d{\cal W}^2}{da} \label{bps} .
\end{eqnarray}
In order to take  advantage of the results already obtained in the case of fluids, we have substituted  ${\cal W}(\s)$  and $V(\s)$  by ${\cal W}(a)=\sqrt{\frac{2}{3} \rho}$  and $V(a)$,  taking into account the explicit expression of  $a(\s)$. The corresponding scale factor self-duality transformations (\ref{homog}) now read
\begin{eqnarray}
&&{\cal  W}(\Omega a)=\Omega^{-1}{\cal W}(a), \quad   \til {\s} (\s)=\pm \int{\sqrt{\frac{2V-{\cal W}^2}{3{\cal W}^2-2V} }}d\s +\s_0 ,\nonumber\\
&& V(\Omega a)= - \Omega^{-2} \Big (V(a) -2 {\cal W}^2(a)\Big), \quad \Omega\equiv\frac{c_0^2}{a^2}. \label{bpsdual}
\end{eqnarray}
When applied to the  example of the modified Chaplygin gas model (\ref{rChap}) with $k=0$, the above equations give
\begin{eqnarray}
{\cal W}(\s)=\sqrt{\frac{2}{3}\rho_{dS}}\left(\cosh\left(\frac{\delta}{\sqrt{2}}\kappa(\sigma-\sigma_0)\right)\right)^{\frac{1}{\delta}},\quad
a(\s)= c_0 \left(\sinh\left(\frac{\delta}{\sqrt{2}}\kappa(\sigma-\sigma_0)\right)\right)^{-\frac{1}{2\delta}} \label{arho}
\end{eqnarray}
These results are sufficient to reproduce the self-dual potential:
\begin{eqnarray} 
V_{sd}(\sigma)=\frac{2}{\kappa^2 L^2_{dS}}\Bigg\{\left[\cosh^2\left(\frac{\delta}{\sqrt{2}}\kappa(\sigma-\sigma_0)\right)\right]^{\frac{1}{\delta}}+
2 \left[\cosh^2\left(\frac{\delta}{\sqrt{2}}\kappa(\sigma-\sigma_0)\right)\right]^{\frac{1-\delta}{\delta}}\Bigg\},\label{poten}
\end{eqnarray}
and to derive the corresponding scalar field duality transformations  as well:
\br 
& \sinh\left(\frac{\delta}{\sqrt{2}}\kappa(\tilde\sigma-\tilde\sigma_0)\right)\sinh\left(\frac{\delta}{\sqrt{2}}\kappa(\sigma-\sigma_0)\right)=1.\label{stils}
\er
\begin{figure}[ht] 
\centering
\includegraphics[scale=0.8]{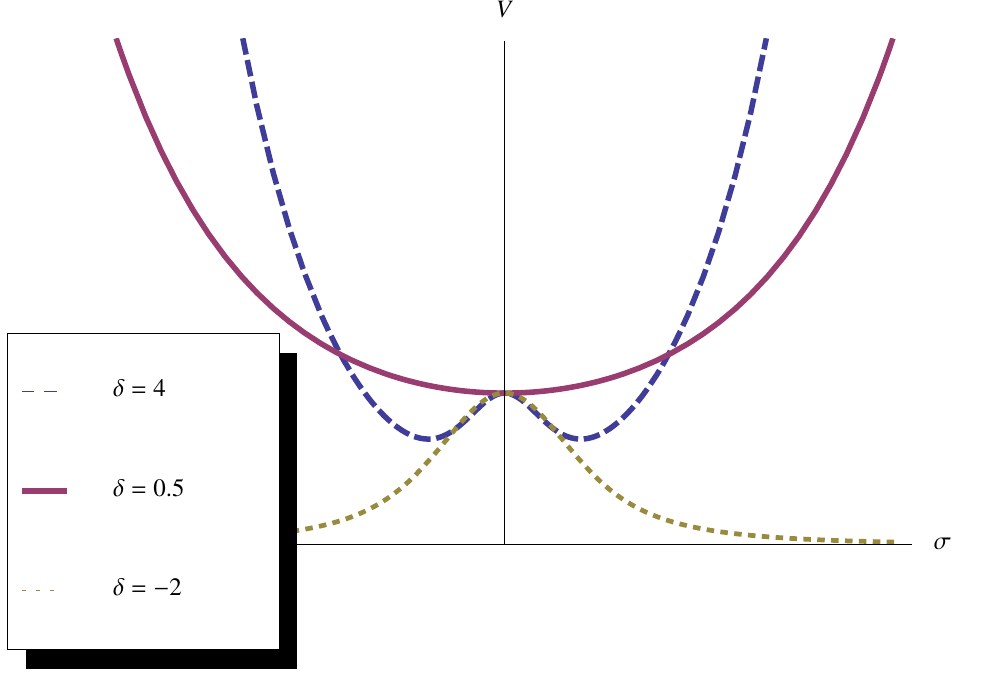}
\caption{Self-dual potential (\ref{poten}) for different values of $\delta$. The de Sitter vacuum sits at $\s_0 = 0$.}
 \label{V_s_PRL2}
\end{figure}
Notice that the above non-linear $\til {\s}(\s)$ duality transformation  is simply related to the scale factor inversions (\ref{sfdro}), i.e.  $\til{a}(\til{\s}) a(\s)=c_0^2$,  and as expected it maps large $\s$ into  small $\til {\s}$, and vice-versa.

The shape of the self-dual potential (\ref{poten}) depends crucially on the parameter $\delta$, as shown in fig.\ref{V_s_PRL2}.
Models with $\delta>3/2$ present a double-well potential, with the maximum centered at $\s = \s_0$ (blue traced line).  For $0< \delta \leq \frac{3}{2}$, the extremum $\s=\s_0$ is a global minimum (solid purple line), while for $\delta < 0$ it becomes a global maximum (dotted yellow line).  
Such differences in shape reflect the allowed  values of  the effective  mass $m^2_{\s}=V''(\s=\s_0)$ of the scalar field $\s$, given by $L^2_{dS} m^2_{\s}=4 \delta \left(\frac{3}{2} - \delta \right)$. 
 As it is seen from eqs. (\ref{roro}) and (\ref{curvtransinvert}), models with $\delta > 1$ present a rather unphysical feature:  the curvature of the universe increases with the deceasing of the matter density. We shall therefore exclude these  models  from the subsequent discussions.

\section{UV/IR symmetric cosmological models} \label{SectUVIR}

\begin{figure}[ht] 
\centering
\includegraphics[scale=0.7]{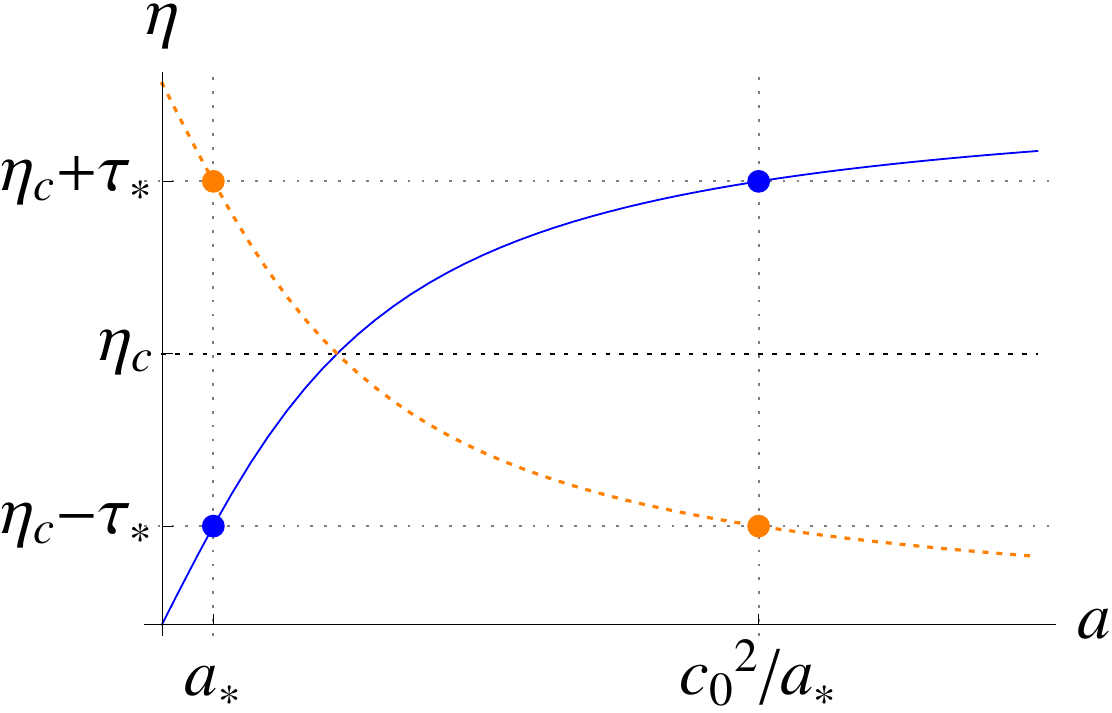}
\caption{Geometric nature of self-duality. Blue curve is $\eta(a/c_0)$, dotted orange curve is $\eta(c_0/a)$. The pairs of blue (and orange) points are related by a scale factor inversion.}
 \label{geometricselfdual}
\end{figure}

The eventual applications of  the simplest family of self-dual models (\ref{poten}) (representing the modified Chaplygin gas (\ref{rChap}), with $k=0$) in the description of the universe evolution require further investigation of the properties of their FRW's solutions in order to establish for which values of the parameters $\delta$ and $L_{dS}$ they might be in agreement with the predictions of selected inflationary and/or $\Lambda$CDM-type models. The simple form of their  energy density (\ref{rChap})
allows to derive the exact solutions $\eta(a)$:
\begin{eqnarray}
 \eta\left(\frac{a}{c_0}\right) = \frac{a L_{dS}}{c^2_0}  \ _{2}F_1\left(\frac{1}{2\delta},\frac{1}{4\delta};1+\frac{1}{4\delta};- \left(\frac{a}{c_0}\right)^{4\delta}\right),\quad  
c_0^{4\delta}= \frac{D}{B}=D\left(\frac{\kappa L_{dS}}{\sqrt{6}}\right)^{2\delta},
\label{solu}
\end{eqnarray}
by direct integration of the first of eqs.(\ref{frw}). 
Their self-duality can be  verified by taking into account the following identity between  hypergeometric functions:
\br
 \eta(a/c_0)=-\eta(c_0/a)+2\eta_c,\quad\quad \eta_c=\frac{L_{dS}}{8\delta c_0}\frac{\left[\Gamma\left(\frac{1}{4\delta}\right)\right]^2}{\Gamma\left(\frac{1}{2\delta}\right)}.\label{etadual}
\er
%
\begin{figure}[ht] 
\centering
\subfigure[]{
\includegraphics[scale=0.5]{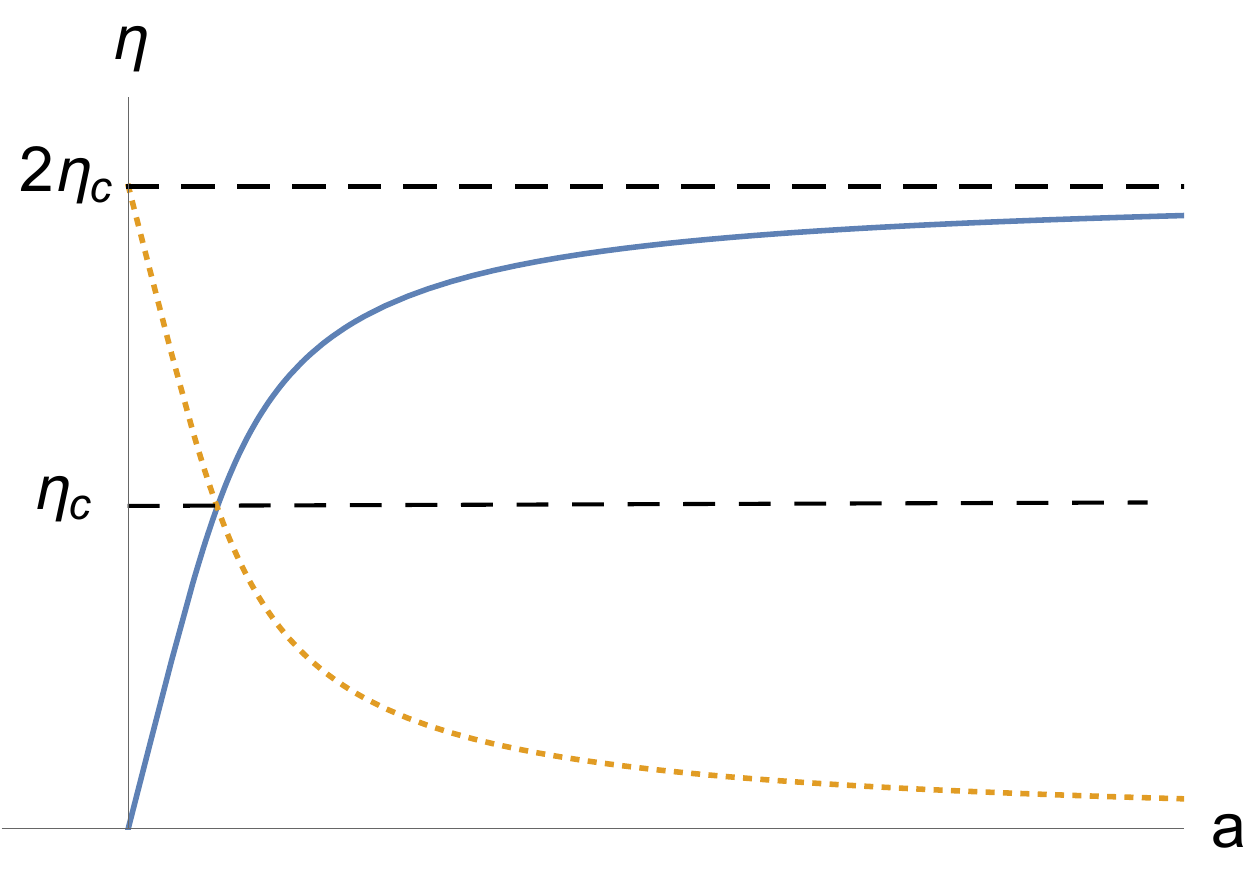} \label{eta_1}
}
\subfigure[]{
\includegraphics[scale=0.5]{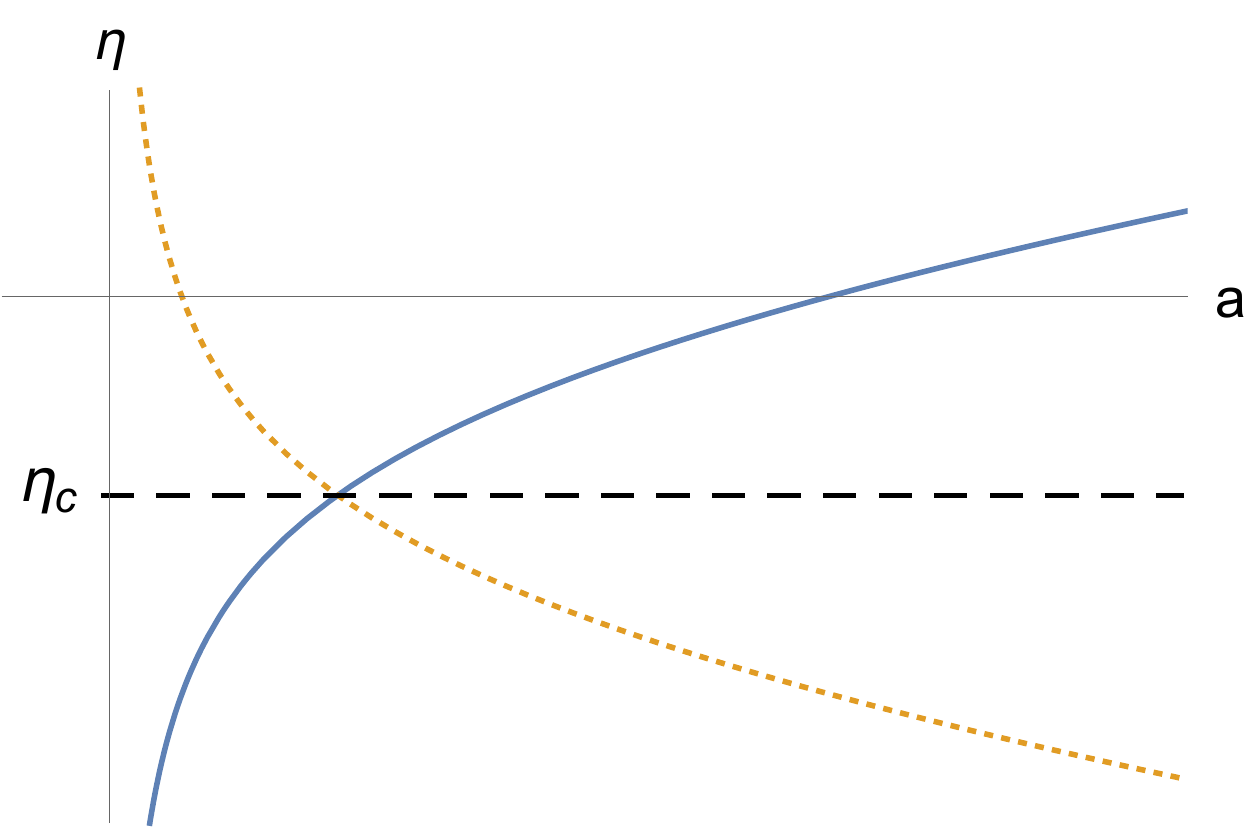} \label{eta_2}
}
\caption{Conformal time as a function of scale factor, given by eq. (\ref{solu}).
 Blue lines represent $\eta(a / c_0)$ and dotted orange lines, $\eta(c_0 / a)$.
 (a) $\delta > 0$; (b) $\delta < 0$.}
 \label{etaaab}
\end{figure}
%
It can also be easily seen as the symmetry between the curves $\eta(a/c_0)$ and $\eta(c_0 / a)$ under a mirror reflection over the line $\eta_c$, as shown in fig.\ref{etaaab}. The geometric meaning of the self-duality condition, expressed in the first of the functional relations (\ref{homog}),  is illustrated in fig.\ref{geometricselfdual}: the blue curve gives the evolution of an universe as $\eta(a/c_0)$; the pair of  ``dual'' blue points have $\eta$-coordinates equidistant from the fixed point $\eta_c$. The scale-factor self-duality assures that the geometries at any such pair of blue points  are related by a simple scale factor inversion  -- as it can be clearly seen with the help of the auxiliary orange curve $\eta(c_0/a)$. 

Although the explicit form of the scale factor $a(\eta)$ can only be obtained in a few particular cases of $\delta=1,1/2,1/4$,  the asymptotic behavior of $a(\eta)$ in the distant future and past of the universe may be read from the asymptotic limits of the hypergeometric function. The particular features depend crucially on the sign of $\delta$, i.e. on the shapes of the qualitatively different matter potentials (see fig.\ref{V_s_PRL2}).

 In case $\delta>0$ -- corresponding to the potential (\ref{poten}) indicated by the purple solid line in fig.\ref{V_s_PRL2} --
 we realize that eq. (\ref{solu}) describes an expanding  universe  with singularity at $\eta=0$ and dominated by radiation, $a^2(\eta)\sim\eta^2$, 
at its initial decelerated phase. It ends, accelerated, as a de Sitter universe  in the ``far future", when $a^2(\eta)\approx L_{dS}^2/(\eta_f-\eta)^2$, which shows that the end of its life is reached at \emph{finite} conformal time  $\eta_f= 2\eta_c$ (see fig.\ref{eta_1}).  Both the surfaces representing the singularity and future infinity are space-like, and therefore these universes present both particle and event horizons.  
If $L_p(\eta)$ is the physical radius of the particle horizon at an instant $\eta$ (necessarily in the decelerated period), then it relates to  the physical radius $L_e(\tilde\eta)$ of  the event horizon at the instant $\tilde\eta = 2\eta_c - \eta$ (necessarily in the accelerated period) through  (cf. eq. (\ref{dhors}))
\br
 L_e(\tilde\eta)=\left(\frac{\til {L}}{L}\right)^{\frac{1}{1-\delta} }L_p(\eta), 
\label{sdhor}
\er
where $L^2(\eta) \equiv 12/ R(\eta)$ and $\tilde L^2 \equiv 12 / R(\tilde \eta)$ are the Ricci curvature radius of space-time at each instant. 
The existence of both types of horizon was to be expected: as discussed in Sect.\ref{SectCtSFD},  duality maps particle into event horizons and vice-versa, therefore a self-dual universe must possess either both types of horizon or none.

The solutions with $\delta<0$ -- corresponding to the yellow dotted line in fig.\ref{V_s_PRL2} -- manifest quite different features, with monotonically decreasing and \emph{bounded} density, $\rho_{dS} > \rho > 0$, and curvature, $R_{dS}> R > 0$, as it is clear form the fact that $V(\s)$ is always finite. Now, $\eta$ is defined on the entire real line, and thus we have an \emph{eternally} expanding universe (see fig.\ref{eta_2}); here, past and future infinity are null hypersurfaces and there are neither particle nor event horizons, nor singularities. As one can see by taking $a\rightarrow0$ in eq. (\ref{solu}), at the remote past ($\eta\rightarrow -\infty$) the space is asymptotically de Sitter, and is accelerated up to the moment $\eta_c$, when the decelerated phase begins. For large and positive $\eta$, in the far future, radiation dominates, with $a(\eta)^2\sim\eta^2$.

Let us consider in more detail the  $\delta = \frac{1}{2}$  example, a self-dual fluid whose energy density 
\br
\rho(a)= \left( B + D \, a^{-2} \right)^2 = \Omega_\Lambda + \frac{\Omega_{rad}}{a^4} + \frac{\Omega_{c}}{a^2}  \label{delta1meiodens}
\er
is a sum of three interacting fluids:  a cosmological constant with $\Omega_\Lambda = B^2$, radiation with $\Omega_{rad} = D^2$, and a cosmic string gas with $\Omega_{c} = 2 BD$.%
\footnote{See also ref. \cite{Lu:2013una} for an interpretation of this model as a unification of dark energy and `dark radiation', as well as for observational constraints.} 
 The EoS is $p=\frac{1}{3} \rho -\frac{4}{3} \sqrt{\Omega_\Lambda \rho}$. A particularly simple form of the $\eta(a)$ solution (\ref{solu})  then provides an explicit expression for the scale factor $a(\eta)$:
\br
a(\eta) = c_0 \tan \left( \frac{c_0}{L_{dS}} \eta \right) ,\quad  0< \eta < 2\eta_c = \frac{L_{dS}}{c_0} \times \pi/2,\quad \frac{c_0}{ L_{dS}} = \kappa \sqrt{\frac{\Omega_c}{12}},
\er
The behavior near the singularity is readily seen by taking $\eta \ll 1$, so $a \sim \eta$, while the far future of the universe is attained in the limit $a \to \infty$, corresponding to the finite value of conformal time
$\eta \to \eta_f = \frac{L_{dS}}{c_0} \times \pi / 2 = 2 \eta_c$.  The simple form of  $a(\eta)$  opens the opportunity  to write this solution as a function of \emph{cosmic} time as well:
\br
a(t)=c_0\sqrt{e^{2L_{dS} t / c_0^2}-1},\quad \quad t = \int a(\eta) d\eta = - \frac{c_0^2}{L_{dS}} \log\left[ \cos \left( \frac{L_{dS}}{c_0} \eta \right) \right] . \nonumber
\er
We can also find the explicit transformation between the dual cosmic times.
After calculating  $\tilde t = \int \tilde a(\tilde\eta) d \tilde \eta = -\int \frac{c_0^2}{a(\eta)} d\eta$, 
we easily find that $\tilde{t}(t) $ is given implicitly by  
\begin{eqnarray}
 e^{- 2 L_{dS} \tilde t / c_0^2} =1 - e^{- 2L_{dS} t / c_0^2}.
   \label{costime}
\end{eqnarray}
We see that the conformal time scale factor duality (\ref{tau}) \emph{can} be realized  as a cosmic time scale factor inversion, $\til {a}(\tilde t)=\frac{c^2_0}{a(t)}$, but with the rather nontrivial cosmic  time transformation above.\footnote{To be compared with the standard cosmic time scale factor duality (see eq. (\ref{chimentdual})), in which $\tilde {a}(\pm t)=\frac{c^2_0}{a(t)}$, i.e. the cosmic time transformation is trivial: $\tilde t=\pm t$.}

\vspace{0.5cm}

The methods developed in the present paper were designed  for the description of selected UV/IR symmetric cosmological models , representing two phases  of decelerated and accelerated  expansion.  The  above analysis of the FRW solutions of the simplest self-dual models (\ref{poten}), however, indicates two distinct  physical regimes that can be asymptotically  described by them:

\emph{Post-inflationary cosmology}: The models with $0<\delta \leq 1$ (purple solid line in fig.\ref{V_s_PRL2})  possess features specific for the  standard model post-inflationary cosmology. We have already seen in eq. (\ref{delta1meiodens}) that the case $\delta = \frac{1}{2}$ corresponds to a  combination of radiation, dark energy and a string gas, but without dust. For $\delta = \frac{1}{4}$ we now have all the elements of the self-dual version of $\Lambda$CDM shown in eq. (\ref{lcdm}):
\br
 \rho=\left(B + D a^{-1} \right)^4 = \Omega_{\Lambda}+\frac{\Omega_{rad}}{a^4}+\frac{\Omega_{dw}}{a}+\frac{\Omega_{dust}}{a^3}+\frac{\Omega_{c}}{a^2} , \nonumber
\er
with a specific relation between the critical densities  
$$\Omega _{dust}=4BD^3,\quad \Omega _{dw}=4DB^3,  \quad \Omega _{c}=6B^2D^2 ,\quad \Omega_{rad} =D^4,\quad \Omega_{\Lambda}=B^4.$$
Modifications of $\Lambda$CDM can also be found. For example, the self-dual potential with $\delta = 3/4$  manifests many of the properties of the  Chaplygin-like \emph{quintessence} models \cite{Kamenshchik:2001ec, Debnath:2004cq, Benaoum:hh, Chimento:kq}. The energy density,  which at early times is approximated by radiation, $\rho \approx \Omega_{rad}/a^4$, at relatively late times gets a contribution from the cosmological constant and dust, $\rho \approx \rho_{dS} +\Omega_{dm}/a^3$.

\emph{Hilltop  inflation}: On the other hand, the  models with $\delta<0$ (yellow dotted line in fig.\ref{V_s_PRL2})  provide  a realization of  \emph{hilltop  inflation} \cite{Boubekeur:2005sw}. The self-dual potential (\ref{poten}), nearby the ``unstable'' de Sitter vacuum at $\s=0$, take a form  $V(\s) \approx 1 - |\delta| \s^2$. Inflation takes place as the field rolls down this maximum, and ends when the field reaches the Planck scale, with $\s_{end} \sim m_{Pl} \equiv 1$. We may then calculate the number of e-foldings before the end of inflation   
$$N(\s) = - \frac{1}{2|\delta|} \int_1^\s \frac{d\s'}{\s'} , $$
and parametrize the field as $\s = e^{- 2|\delta| N}$. The spectral index $n_s$ and the tensor-to-scalar ratio $r$, are then
\br
n_s = 1 - 4 |\delta| - 12 \delta^2 e^{-4 |\delta| N} , \quad r = 32 \delta^2 e^{-8 |\delta| N} . \nonumber
\er
Setting $N = 60$ at the moment of horizon crossing and taking  $\delta \approx -0.075$, we find
$n_s \approx 1+4\delta\approx 0.97$, and $r \sim 10^{-5}$,  which are in good agreement with the recent Planck2015 data \cite{2015arXiv150201589P,2015arXiv150202114P}.  

As we have mentioned, more realistic inflationary and  quintessence models (or even generalizations of $\Lambda$CDM) can be constructed  when self-dual fluids with a more general EoS or with more components are considered, as in the examples given by eqs. (\ref{lcdm}), (\ref{twogen}) and  (\ref{MultiChap}). Although they have not an explicit scalar field representation and require specific relations between the values of the critical densities of the constituent fluids, their  advantage is to offer an equivalent description of the early time short distances physics  in terms of the late time large distances one, guaranteed by scale-factor self-duality.

\section{Concluding remarks}


The most intriguing result of our study of the conformal time scale factor self-duality transformations is their  ``almost Weyl''  form, seen in eqs. (\ref{homog}) and (\ref{bpsdual}),  as rescalings of the metric, superpotential, etc. -- but with the exception of the matter field transformations (\ref{stils}). The question arrises of whether one can find  an equivalent form of the considered models where the new matter field obeys a standard Weyl transformation.
  We begin with the encouraging observation that by a simple  change of the field variable $\phi^2= \Phi \bar {\Phi}=6 \tanh^2(\delta \s )$ (with $\kappa^2=2$) our self-dual models (\ref{poten})  are transformed into  $SU(1,1)/U(1)$ K\"ahler sigma models,\footnote{With the usual  Kahler metric  $g^{\Phi \bar {\Phi}}= \partial_{\Phi} \partial_{\bar {\Phi}} K$ of constant curvature $R_{K} =-4 \delta^2$.} with gauge fixed $U(1)$ symmetry $\Phi=\bar{\Phi}=\phi$ and minimally coupled to Einstein gravity. Indeed such an embedding into ${\cal N} =1$ supergravity with one chiral matter supermultiplet  is  quite natural  and expected due to its well known role in the construction of both  quintessence \cite{Barreiro:2000mi,Copeland:2006if,Kamenshchik:2001ec, Binetruy:1999qf,Brax:1999ee} and  inflationary models \cite{Kallosh:2010uo,Kallosh:2002ph,Kachru:2003jt, Kallosh:2014dw,Ferrara:2013fv,Ferrara:2010zp}. The new feature, however, consists in a very special relation between the K\"ahler potential $K(\Phi,\bar{\Phi})$ and the superpotential $|{\cal Z}|={\cal W}(\phi)=e^{\frac{K}{2}} |W|$,
 \begin{eqnarray} 
& K=-\frac{1}{2\delta^2} \ln\Big(1-\frac{1}{6} \Phi \bar{\Phi}\Big) ,\quad |{\cal Z}|=\frac{\sqrt{2}}{L_{dS}} e^{\delta K},
\label{kahler}
\end{eqnarray}
where $W(\Phi )$ denotes the  holomorphic superpotential  of  ${\cal N} =1$ Poincar\'e supergravity \cite{Kallosh:2014dw,Cvetic:1995jo,Cvetic:1997fx,Townsend:2007ta,Skenderis:2007yb}.  It is worthwhile to also  mention that
 the inflationary $\alpha$-attractors \cite{Kallosh:2013jk,Kallosh:2014rz}, obtained within the  superconformal supergravity approach \cite{Ferrara:2013fv,Ferrara:2010zp},
for a special choice of the inflaton  potential:
$$f_{sd}^2\left(\frac{\phi}{\sqrt{6}}\right)=V_{sd}(\phi)=\frac{1}{9} \Big(\frac{6B}{6-\phi^2}\Big)^{\frac{1}{\delta}} (9-\phi^2), \ \alpha=\frac{1}{6\delta^2}$$ 
turns out  to coincide with  our model (\ref{poten}).
There exists however an important difference   between the well known ``boundary'' conditions for the $\alpha$-attractor potentials and those of our self-dual example, $f^2_{sd}(0)\neq 0$  and $f^2_{sd}(\pm1)=\infty$ (for $\delta>0$), while for the attractors of ref. \cite{Kallosh:2013jk,Kallosh:2014rz} $f^2_{att}(0)=0$  and $f^2_{att}(1)={\rm{const.}}<\infty$. 
 As a consequence,  the vacua at the extremum $\phi=0$ represent  different  geometries:  Minkowski for attractors and de Sitter for the self-dual potential (\ref{poten}),  thus indicating that the supersymmetry  is also broken on the latter.

With such a K\"ahler sigma model representation of the  self-dual model (\ref{poten}), we may realize the  conformal time SFD  transformations (\ref{sfsd1}) as a particular $Z_2$ Weyl transformation,  now eqs. (\ref{sfdro}), (\ref{homog}) and (\ref{stils}) assuming a simple ``linear form'':
\begin{eqnarray} 
\til{\phi}(\bar{\eta})=\Omega^{-\delta} \phi(\eta), \quad\quad \til{a} (\bar{\eta})=\Omega a(\eta), \quad\quad \Omega(\phi)=\Big(\frac{\phi^2}{6-\phi^2}\Big)^{\frac{1}{2\delta}}. 
\label{weyl}
\end{eqnarray} 
 We should note these relations are valid \emph{on-shell} -- the knowledge of the explicit form of the scale factor $a(\s)$ as a function of the scalar field is needed in order to derive the $\til {\s}(\s)$ transformations (\ref{stils}), as well as  their Weyl counterparts. 
The above form of the considered  $Z_2\times Z_2$ symmetry also suggests an  enlightening  interpretation of the requirement for scale factor self-duality.  It can be considered as an effective rescaling of the field $\phi$ and of the metric by a dynamically determined dilatation factor $\Omega(a_*)>1$ at fixed time $\eta_*<\eta_c$,  which when combined  with the time reflection  $\tilde{\eta}_*=2\eta_c -\eta_*$  ensures the equivalence of the short-distance physics at  $\eta_*$ with the large-distance one  at  the  ``dual''  instant $\tilde{\eta}_*$.

Our investigation of the self-duality  aspects of the  \emph{conformal time scale factor inversions}  has  revealed they have an interesting new face as a UV/IR symmetry of a family of cosmological models manifesting  two periods of expansion.  
As discussed in Sect.\ref{SectCtSFD}, the problems we have addressed, although bearing some similarities, are not identical to the original  dilaton gravity applications  of  the scale factor duality \cite{Veneziano91scalefactor,Gasperini:2003sh}. It is nevertheless worthwhile to mention that some of its most notable  pre-big-bang features \emph{can} be realized in our Einstein gravity cosmological models -- one can construct a class of pre-big-bang-like self-dual solutions, despite the absence of the dilaton. 
%
%
%
 This is achieved by taking an expanding solution of model (\ref{rChap}) for $\delta>0$ (the blue curve in fig.\ref{eta_1}), and gluing it, at $\eta=0$, 
with its  contracting  image (the dotted orange line in fig. \ref{eta_1}) obtained  by the SFD inversion (\ref{sfdro}) together with a particular time \emph{translation}, $\bar{\eta}=\eta-2\eta_c$. The result is a past-extended solution with an initial and a final accelerated periods (asymptotically de Sitter in the far future and past, $\eta \to \pm 2 \eta_c$) separated by a decelerated phase in the middle of which (at $\eta=0$) there is a big-bang singularity.
%
It is  then natural to expect that  the cosmological  models we have studied  might be identified as  an Einstein frame version of certain pure  dilaton gravity models (without any matter fluids), but with non-trivial potentials fixed by the self-duality requirement. This also suggests that  some of
 the established  self-duality  properties   may be observed in their Jordan frame  realizations.

\bibliographystyle{JHEP}

\bibliography{References}

\end{document}